\documentclass[aps,prd,twocolumn,groupedaddress]{revtex4}  
\usepackage{graphicx}  
\usepackage{dcolumn}   
\usepackage{bm}        
\usepackage{amssymb}   
\usepackage{braket} 

\newcommand{\eqref}[1]{{(\ref{#1})}}

\newcommand{\be}{\begin{equation}}
\newcommand{\ee}{\end{equation}}
\newcommand{\bea}{\begin{eqnarray}}
\newcommand{\eea}{\end{eqnarray}}

\newcommand{\ti}{\times}

\newcommand{\mc}{\mathcal}

\newcommand{\beqa}{\begin{eqnarray}}
\newcommand{\eeqa}{\end{eqnarray}}

\newcommand{\beq}{\begin{equation}}
\newcommand{\eeq}{\end{equation}}

\newcommand{\axion}{\Ket{a}}

\begin{document}






\title{A 3.55 keV Photon Line and its Morphology from a 3.55 keV ALP Line}




\author{Michele Cicoli}\email{mcicoli@ictp.it}\affiliation{Dipartimento di Fisica ed Astronomia, Universit\`a di Bologna, via Irnerio 46, 40126 Bologna, Italy \\ INFN, Sezione di Bologna, 40126 Bologna, Italy \\
Abdus Salam ICTP, Strada Costiera 11, Trieste 34014, Italy}
\author{Joseph P.~Conlon}\email{j.conlon1@physics.ox.ac.uk} \affiliation{Rudolph Peierls Centre for Theoretical Physics, University of Oxford, 1 Keble Road, OX1 3NP, Oxford, United Kingdom}
\author{M.C.~David Marsh}\email{david.marsh1@physics.ox.ac.uk}\affiliation{Rudolph Peierls Centre for Theoretical Physics, University of Oxford, 1 Keble Road, OX1 3NP, Oxford, United Kingdom}
\author{Markus Rummel}\email{markus.rummel@physics.ox.ac.uk}\affiliation{Rudolph Peierls Centre for Theoretical Physics, University of Oxford, 1 Keble Road, OX1 3NP, Oxford, United Kingdom}
\date{\today}
\begin{abstract}
Galaxy clusters can efficiently convert axion-like particles (ALPs) to photons.
We propose that the recently claimed detection of a 3.55--3.57 keV line in the stacked spectra of a large number of galaxy clusters and the
Andromeda galaxy may originate from the decay of either a scalar or fermionic $7.1$ keV dark matter species into an axion-like particle (ALP) of
mass $m_{a} \lesssim 6\cdot 10^{-11}~{\rm eV}$, which subsequently converts to a photon in the cluster magnetic field.
In contrast to models in which the photon line arises directly from dark matter decay or annihilation,
 this can explain the anomalous line strength in the Perseus cluster.
 As  axion-photon conversion scales as $B^2$ and cool core clusters have high central magnetic fields, this
 model can also explain the observed peaking of the line emission in the cool cores of the Perseus, Ophiuchus and Centaurus clusters,
 as opposed to the much larger dark matter halos. We describe distinctive predictions of this scenario for future observations.
\end{abstract}


\maketitle

Recently two analyses have appeared which claim the possible existence of a photon line at $E \sim 3.55~\hbox{keV}$, based on stacked data from galaxy
clusters and also from the Andromeda galaxy \cite{14022301, 14024119}.
 In \cite{14022301}, the line was found in the stacked spectrum of 73 galaxy clusters independently in observations with the XMM-Newton PN and MOS instruments to a high (4-5$\sigma$) statistical significance.  
 After dividing the full sample into the three subsamples (Perseus, Coma+Ophiuchus+Centaurus, and all others), the signal was found in all three subsamples by the MOS instrument, by PN in the `all others' subsample, and also with \emph{Chandra} observations of the Perseus cluster. In reference  \cite{14024119}, a similar line was found at 3.5 keV in the XMM-Newton data of the Andromeda galaxy and Perseus cluster.

Perhaps the most exciting interpretation of this line is as originating from dark matter decay to produce photons.
This has been discussed in the context of sterile neutrinos, axion-like dark matter, axinos and excited states of dark matter \cite{Ishida:2014dlp, Finkbeiner:2014sja, Higaki:2014zua, Jaeckel:2014qea, Lee:2014xua, Abazajian:2014gza, Krall:2014dba, Aisati:2014nda, Kong:2014gea, Hamaguchi:2014sea, Frandsen:2014lfa, 14031710, 14031733, 14031782}.


However, there are aspects of this potential signal that are inconsistent with the interpretation as dark matter decaying directly to photons.
\begin{enumerate}
\item
The signal found from the Perseus cluster is much stronger than the signal found from other galaxy clusters.
A line arising from dark matter decay to photons can only trace the quantity of dark matter in each cluster.
This makes a clear prediction for the relative magnitude of signal from each cluster.
However, the effective inferred dark matter decay rate from Perseus is a factor four to eight greater than for the stacked sample of clusters (depending on whether
the centra $1^{'}$ is included or not, and depending on whether XMM or \emph{Chandra} data is used).
\item
The effective inferred decay rate for the nearby bright clusters Coma, Ophiucus and Centaurus is also
at mild tension ($1.8\sigma$) with the effective inferred decay rate
for the stacked sample of more distant clusters.
\item
Perseus is the archetypal cool core cluster, and in Perseus both XMM and \emph{Chandra} observations show
a large fraction of the signal arising from the cool core at the very center of the cluster.
In the analysis of \cite{14022301}, excising the central arcminute  --- a radius of approximately 20 kpc --- removed
around half the inferred signal for the XMM MOS detector. The cool core arises from the collisional cooling of the central dense ICM gas and is on far smaller scales
than that of the dark matter halo.
As the physics that leads to the formation of the cool core is entirely astrophysical, any signal from direct dark matter decay to photons should not peak
on these scales.
\item
The line signal from a stacked sample of three nearby bright clusters (Coma, Ophiuchus and Centaurus) is also dominated by emission from the cool
cores of Ophiuchus and Centaurus (note that Coma is not a relaxed cluster and does not have a cool core). As above,
the extent of the cool core is set by astrophysical processes and will not trace the dark matter distribution.
\end{enumerate}
Such features, if persisting in a fuller subsequent analysis, would be inconsistent with an interpretation of the line in terms
of dark matter decaying to photons.

In this paper we propose a model which can reconcile these features with a dark matter origin of the 3.55 keV line. Our proposal is
that the dark matter decay generates a monochromatic 3.55 keV line for an axion-like particle (ALP), which then converts into a
3.55 keV photon in the magnetic field of the galaxy cluster.

The existence of axion-like particles is theoretically well motivated. The QCD axion is by far the most plausible solution
to the strong CP problem of the Standard Model. Axion-like particles frequently arise in compactifications of string theory
\cite{hepth0602233, SvrcekWitten, 12060819} and there is an active experimental program searching for their existence \cite{12105081, 13025647, 14013233}.
The observability of axion-like particles is set by their coupling to photons,
$$
\frac{a}{M} {\bf E} \cdot {\bf B} \, ,
$$
which implies that ALPs convert to photons in background magnetic fields. For massless ALPs, $M$ is bounded by $M \gtrsim 10^{11} \hbox{GeV}$.

The intracluster medium of galaxy clusters is pervaded by large-scale turbulent magnetic fields. The existence of these magnetic fields is established
from observation of synchrotron emission from clusters in the form of radio halos, minihalos or relics, and from Faraday Rotation Measures of background radio sources
with the cluster as a Faraday screen. These measurements imply that cluster magnetic fields are generally $ B \sim \mc{O}(\mu G)$, with larger values of
$B \sim \mc{O}( 10~\mu G)$ found near the center of cool core clusters. The magnetic field is multiscale, with typical
coherence lengths $L \sim 1 - 10~\hbox{kpc}$ which can extend to $L \sim 100 ~\hbox{kpc}$.

This intracluster medium can in fact be an efficient convertor of ALPs to photons (e.g. see \cite{Burrage, CAB, 13123947}).
For $M \sim 10^{11} ~\hbox{GeV}$,
a massless ALP with X-ray energies will have a conversion probability that is at or close to the saturation level
of $\langle P_{a \to \gamma} \rangle = 1/3$, although with stochastic variations that depend on the line-of-sight realisation of the magnetic field.
Any significant source of X-ray ALPs in a cluster can then generate an appreciable source of X-ray photons.

The conversion of X-ray ALPs to photons in the Coma cluster has been studied in detail in \cite{13123947}. Many string theory scenarios
of the early universe predict a dark radiation cosmic axion background in the $0.1 - 1 ~\hbox{keV}$ waveband, originating from moduli
decays in the early universe. Such axions can convert
to photons in the cluster magnetic field, generating a broad excess X-ray flux which may explain the long-standing excess
in soft X-rays from many galaxy clusters \cite{Lieu1996, Bonamente2002, Durret2008}, which is particularly well established for the case of Coma
\cite{Lieu1996B, astroph9911001, Bonamente2002, Bulbul2009}.

In \cite{13123947}, axions were propagated through a full $2000^3$ simulation of the magnetic field of the Coma cluster using
the magnetic field parameters determined in \cite{Bonafede} as a best fit to Faraday Rotation Measures.
For this case of the Coma cluster, it was found that for an axion-photon coupling $M \sim 10^{13} ~\hbox{GeV}$,
the central conversion probability of a $3.55 ~\hbox{keV}$ axion was $P_{a \to \gamma} \sim 10^{-3}$. This probability scales
as $\left( \frac{10^{13} \text{GeV}}{M} \right)^2$.

As described in the original papers \cite{14022301, 14024119}, if the 3.55 keV line is produced by dark matter decaying directly to photons,
the dark matter lifetime is $\tau \sim 10^{28} ~\hbox{s}$. As the age of the universe is $\tau \sim 4 \ti 10^{17} ~\hbox{s}$, it is clear
that there is significant room for a shorter lifetime for decay to axions $ \tau \ll 10^{28} ~\hbox{s}$ balanced by a conversion
probability $P_{a \to \gamma} \ll 1$. We impose a conservative value $\tau > 2\cdot 10^{19} ~\hbox{s}$ for the dark matter lifetime \cite{Gong:2008gi}.

\subsection{Models for Axion Production}

A monochromatic ALP can be produced by the decay of either scalar or fermionic dark matter.
For the scalar case, an example is moduli dark matter, which can decay to axions via the kinetic coupling
$$
\frac{ \Phi}{\Lambda} \partial_{\mu} a \partial^{\mu} a \, .
$$
This coupling has been considered in the context of dark radiation in \cite{12083562, 12083563, 13047987, 13054128}. This induces decays of moduli to ALPs  with  a decay rate of
\be
\Gamma_{\Phi} =  \frac{1}{32 \pi} \frac{m_{\Phi}^3}{\Lambda^2} \, .
\ee
The corresponding  lifetime is then
\be
\tau_{\Phi} = \left( \frac{7.1 ~\hbox{keV}}{m_{\Phi}} \right)^3 \left( \frac{\Lambda}{10^{17} ~\hbox{GeV}} \right)^2 1.85 \ti 10^{27} ~\hbox{s} \, .
\ee
An explicit string model with stabilised moduli which features a very light modulus
with these properties has been derived in \cite{Cicoli:2008va} in the context of type IIB LARGE
Volume Scenarios. The model described in \cite{Cicoli:2008va} is characterised by the presence
of two very light moduli $\phi_1$ and $\phi_2$ with masses:
\be
m_{\phi_1}\simeq M_P \epsilon^{5/3}\ll m_{\phi_1}\simeq M_P \epsilon^{3/2}
\ee
with $\epsilon = \frac{m_{3/2}}{M_P}\ll 1$. TeV-scale supersymmetry in this model requires $\epsilon\simeq 10^{-14}$,
leading to the prediction: $m_{\phi_1}\simeq \mathcal{O}(1)$ MeV and $m_{\phi_2}\simeq\mathcal{O}(10)$ keV.
Similarly to what has been recently proposed in \cite{14031733}, these light moduli could not create any cosmological problem in the presence of a
primordial mechanism that suppresses their misalignment during inflation \cite{Linde:1996cx}.
In particular, $\phi_2$ would have a life time of order $\tau\sim 10^{27} ~\hbox{s}$
and could account for most of the dark matter density without overclosing our Universe.

For the fermionic case, a massive fermionic dark matter particle $\psi$ can decay to a 
 fermion $\chi$ and an ALP as $\psi \to \chi a$ via the coupling
\be
\frac{\partial_{\mu} a}{\Lambda} \bar{\psi} \gamma^{\mu} \gamma^5 \chi \, .
\ee
This generates a tree-level decay $\psi \to \chi a$, with a rate of
\be
\Gamma_{\psi \to \chi a} = \frac{1}{16 \pi} \frac{(m_{\psi}^2- m_{\chi}^2)^3}{m_{\psi}^3 \Lambda^2} \, ,
\ee
which for $m_{\chi} \ll m_{\psi}$ corresponds to a lifetime of
\be
\tau_{\psi} = \left( \frac{7.1 ~\hbox{keV}}{m_{\psi}} \right)^3 \left( \frac{\Lambda}{10^{17} ~\hbox{GeV}} \right)^2 0.92 \ti 10^{27} ~\hbox{s} \, .
\ee
While this has been written for general fermionic dark matter $\psi$, we see no reason this may not also exist for the particular case of a
massive sterile neutrino (as an additional decay channel to $\nu \gamma$).

Although axions are generally weakly coupled to matter,
when dealing with lifetimes $\tau > 10^{20}~\hbox{s}$, there is no good reason to neglect axionic decay channels.
Given the often-considered range for the QCD axion of
$f_a \sim 10^9 - 10^{12} ~\hbox{GeV}$,
we see that the axionic coupling constants considered above are entirely reasonable from a particle physics perspective.

\subsection{Axion-photon conversion}

Once an axion is produced, axion to photon conversion occurs via the operator
\be
\label{eq:MixOp}
\frac{a}{M} {\bf E} \cdot {\bf B} \, .
\ee
It follows that the morphology and strength of an observed photon line signal is set by the magnetic field environment.
At the simplest level, the signal scales as the square of the magnetic field, although as we discuss below the magnetic field coherence length
and the electron density also play significant roles.
It is this that both distinguishes the predictions of our model from the many variants of dark matter decaying directly to photons, and
also allows it to explain aspects of the data that are inconsistent with conventional explanations.

After including the operator \eqref{eq:MixOp}, the linearised equations of motion for  axion-photon modes of energy $\omega$ is given by,
\beq
\label{eq:EqofMotion}
\left(\omega + \left(\begin{array}{c c c}
			\Delta_{\gamma} & \Delta_{F} & \Delta_{\gamma a x} \\
			\Delta_{F} & \Delta_{\gamma} & \Delta_{\gamma a y} \\
			\Delta_{\gamma a x} & \Delta_{\gamma a y} & \Delta_{a}
		   \end{array}\right) - i\partial_z\right)\left(\begin{array}{c}
								\Ket{\gamma_x} \\
								\Ket{\gamma_y} \\
								\axion
							      \end{array}\right)= 0 \, .
\eeq
Here,
$\Delta_{\gamma}  = -\omega_{pl}^2/2\omega$, where
\bea
\omega_{pl} = \left(4 \pi \alpha \frac{n_e}{m_e}\right)^{1/2} =1.2\cdot10^{-12} \sqrt{\frac{n_e}{10^{-3} {\rm cm}^{-3}}}~{\rm eV} \nonumber
\eea
 denotes the plasma frequency of the ICM. Furthermore, $\Delta_a  = -m_a^2/\omega$,  $\Delta_{\gamma a i}  = B_i/2M$, and
$\Delta_F$, which will be unimportant for the subsequent discussion, denotes  the  Faraday rotation of photon polarisation states caused by  the cluster magnetic field.

Equation \eqref{eq:EqofMotion} is easily solved for a single domain of length $L$ with a constant magnetic field.
 Denoting the   component  of the magnetic field transverse to the motion by $B_{\perp}$, the resulting conversion probability is given by \cite{Sikivie, Raffelt},
\be
\label{def} \label{eq:singleDomain}
P_{a \rightarrow \gamma}^{\rm single~domain} = \sin^2(2 \theta) \sin^2\left( \frac{\Delta}{\cos 2 \theta}\right) \, ,
\ee
where  $\tan 2 \theta = \frac{2 B_{\perp} \omega}{M m^2_{eff}}$, $\Delta = \frac{m_{eff}^2 L}{4 \omega}$,  $m^2_{eff} = m_a^2 - \omega_{pl}^2$.

For  ALP masses smaller than the plasma frequency in the cluster, we can write $m_{eff}^2 = - \omega_{pl}^2$, and the angles $\theta$ and $\Delta$  evaluate to
\bea
|\theta| &\approx& \frac{B_{\perp} \omega}{M m^2_{eff}} = 5.0 \cdot 10^{-4} \times \left( \frac{10^{-3} ~\hbox{cm}^{-3}}{n_e} \right)
  \left(\frac{B_{\perp}}{1~{\rm \mu G} }\right)  \nonumber \\
&&\times   \left(\frac{\omega}{3.55~{\rm keV} }\right)  \left(\frac{10^{14}~{\rm GeV} }{M }\right) \, . \label{eq:theta} \\
\Delta & = & 0.015 \times  \left( \frac{n_e}{10^{-3} ~\hbox{cm}^{-3}} \right) \left(\frac{3.55~{\rm keV}}{\omega} \right) \left(\frac{L}{1~{\rm kpc}}\right)\, . \label{eq:Delta}
\eea
In the small-angle approximation $|\theta| \ll1$ and $\Delta \ll1$, the single domain conversion probability is simply given by
\bea
\label{abc}
&&P_{a \rightarrow \gamma}^{\rm single~domain} = \frac{1}{4} \left( \frac{B_{\perp} L}{M}\right)^2  \nonumber \\
&&=  2.3 \cdot 10^{-10}\times \left(\frac{B_{\perp}}{1~{\rm \mu G} }\frac{L}{1~{\rm kpc}}\frac{10^{14}~{\rm GeV} }{M }\right)^2 \, .
\eea

Given a model of the turbulent, multi-scale  magnetic field in galaxy clusters, equation \eqref{eq:EqofMotion} can be solved numerically as was done in \cite{13123947} for the particular case of the Coma cluster.  However, as noted in \cite{13123947}, many of the features of the resulting conversion probabilities can be already understood from the single-domain formula, equation \eqref{eq:singleDomain}.  Thus, we expect that for the purpose of order-of-magnitude estimates and scalings, the conversion probability in a cluster is sufficiently well approximated by
\bea
P_{a \rightarrow \gamma}^{\rm cluster}(\bar B, \bar L) &\approx& \sum_i P_{i, a \rightarrow \gamma}^{\rm single~domain}
\nonumber \\
 &=& \frac{R_{cluster}}{\bar L} P_{ a \rightarrow \gamma}^{\rm single~domain}(\bar B, \bar L) \nonumber \\
 &\rightarrow&  \frac{\bar B^2 L R_{cluster}}{4 M^2} \, ,\label{Pcluster}
\eea
where $R_{cluster}$ is a measure of the size of the cluster and where, in the last line, we have imposed the small $\theta$ and small $\Delta$ approximation.

For axion masses $m_a \gg \omega_{pl}$, the axion to photon conversion probability scale like $P_{a\to \gamma} \sim m_{a}^{-4}$, thus rapidly making the conversion process inefficient. For $\tau_{DM} \gtrsim 2 \cdot 10^{19}$ s, this constrains the ALP mass to $m_{a} \lesssim 6 \cdot 10^{-11}$ eV.




\subsection{An X-ray line from an ALP line}
In our model, the observed photon flux of the 3.55 keV photon line from a source at  red-shift $z$ and luminosity distance $d(z)$ is 
given by
\begin{equation}
 F = \frac{\Gamma_{DM\rightarrow a}}{4\pi d(z)^2}(1+z) \int_V \frac{\rho_{DM}}{m_{DM}} P_{a \rightarrow \gamma}\,dV  \, .\label{FluxALP}
\end{equation}
Let us work out the coupling parameters required for our model. An exact calculation requires the detailed
magnetic field profile of the clusters that contribute to the signal, to determine axion conversion probabilities when propagated through the cluster.
However, as an an approximate guide to parameters, we take those computed for the central region of Coma, for which a
 massless ALP with energy $E_a = 3.55 ~\hbox{keV}$ will convert to a photon with $P^{\rm Coma}_{a \to \gamma} \sim 10^{-3}$ for $M \sim 10^{13} ~\hbox{GeV}$.

Reproducing the 3.55 keV line by direct decays to photons requires a lifetime of $\tau \sim 5 \ti 10^{27} ~\hbox{s}$. We therefore see that
for decays to axions, we require
\be
\tau_{DM\to a} \sim 5 \ti 10^{24} ~\hbox{s} \left( \frac{10^{13} ~\hbox{GeV}}{M} \right)^2.
\ee
Applying a conservative restriction $\tau_{DM} \ge 2\cdot 10^{19} ~\hbox{s}$, we see that the line signal can be reproduced for $M$ as large as $M \sim 5 \ti 10^{15} ~\hbox{GeV}$. Note that even though our model has in principle two free parameters $\tau_{DM\rightarrow a}$ and $M$, observations in terms of photon fluxes only depend on $\tau_{DM\rightarrow a}/M^2$. Hence, as far as Occam's razor is concerned, there is effectively only one free parameter in our model, as there is in dark matter decaying to photons.

Combining the two processes (dark matter decay to axions and axion conversion to photons),
we see that reproducing the observed signal requires
\be
\left( \frac{\Lambda}{2.7 \ti 10^{14} ~\hbox{GeV}} \right) = \left( \frac{2.7 \ti 10^{14} ~\hbox{GeV}}{M} \right).
\ee
This holds for $\Lambda \gtrsim 5 \ti 10^{12} ~\hbox{GeV}$ (to ensure the dark matter lifetime is $\tau_{DM} > 10^{19} ~\hbox{s}$) and $M \gtrsim 10^{11} ~\hbox{GeV}$
(at which point the $a \to \gamma$ conversion probability in clusters saturates at $\langle P_{a \to \gamma} \rangle = 1/3$.
This equation must be read as an approximate relation. The actual values of the axion-photon conversion probabilities will vary from cluster to cluster
depending on the cluster electron density and the magnitude and coherence lengths of the magnetic field in each cluster.

The central magnetic field in the cool core
of the Perseus cluster has been estimated at the relatively large $B \sim 25~\mu$G \cite{astroph0602622}.
Centaurus is also a cool core cluster, with a magnetic field estimated as  $B \sim 11 - 25~\mu$G \cite{07083264}.
We were unable to find an estimate for the central magnetic field for Ophiuchus (another cool core cluster), although
generally Faraday Rotation Measures indicate
$\sim \mc{O}(10)~\mu$G  magnetic fields in the centre of cool core clusters, e.g. for A2199 see \cite{VaccaA2199}.
In contrast estimates of central field strengths for non-cool core clusters give lower values: for
example $B \sim 4.7~\mu$G for Coma \cite{Bonafede}, $B \sim 2 - 2.5~\mu$ G for A2255 \cite{GovoniA2255}
or $B \sim 1~\mu$G for A665 \cite{VaccaA665}.

The central electron density of the Perseus cluster at a radius of $r \sim 10 \hbox{kpc}$ has been measured
as $n_e \sim 3 \ti 10^{-2} ~\hbox{cm}^{-3}$ \cite{ChurazovForman, 08014253}.
It follows that one expects the small angle approximation to be a relatively good approximation throughout the cluster.


\subsection{Perseus}
In this section, we provide a more detailed estimate of the morphology of the 3.5 keV line as arising  from axion-photon conversion of a 3.5 keV ALP line, and compare this estimate to the morphology of a line from decaying dark matter. To this end, we will use a simplistic model of the Perseus cluster magnetic field which we expect to correctly capture our main point: the flux from axion-photon conversion decays much faster with radial distance from the centre of the cluster than what can be expected for decaying dark matter.

The flux from dark matter decaying directly into photons is given by
\begin{equation}
 F_{ DM\to \gamma} = \frac{\Gamma_{DM\rightarrow a}}{4\pi d^2(z)}(1+z) \int_V \frac{\rho_{DM}}{m_{DM}}\,dV\, .\label{FluxDM}
\end{equation}
For the dark matter density  in the Perseus cluster  we take a Navarro-Frenk-White (NFW) profile
\begin{equation}
 \rho_{DM}(r) = \frac{\rho_0}{r/R_s\left(1 + r/R_s\right)^2}\,,
\end{equation}
with $R_s=360$ kpc~\cite{Simi}. This completely specifies the expected flux from dark matter decaying directly into photons.

In contrast, to evaluate the photon flux arising from equation \eqref{FluxALP}, we need to  estimate the conversion probability $P_{a \rightarrow \gamma}$ in the cluster. The electron density in Perseus is well approximated by the double
$\beta$-function \cite{ChurazovForman}
\begin{equation}
n_e(r) = \frac{3.9 \times
10^{-2}\,\text{cm}^{-3}}{\left(1+(r/80\,\text{kpc})^2 \right)^{1.8}} +
\frac{4.05 \times 10^{-3}\,\text{cm}^{-3}}{\left(1+(r/280\,\text{kpc})^2
\right)^{0.87}}\, ,\label{nebeta}
\end{equation}
from which we note that even in the central $r\lesssim100$ kpc cool core region of the cluster, the electron density is no larger than $n_e \approx 4\cdot 10^{-2}$ cm$^{-3}$.  In this region the magnetic field can be expected to be turbulent over $\mc{O}(1)$ kpc scales. It then follows from equation \eqref{eq:Delta} that the angle $\Delta$ is maximised in the central region of the cluster, but  only leaves the small angle regime very close to the centre. For a detailed discussion on the behaviour of the axion-photon conversion probability in close to $\Delta =1$, we refer to \cite{13123947}. Here, we simply note that the small $\Delta$ approximation should be sufficient for our estimate of the conversion probability in both the centre and more remote regions of the cluster.

\begin{figure}[t!]
\centering
\includegraphics[width=\linewidth]{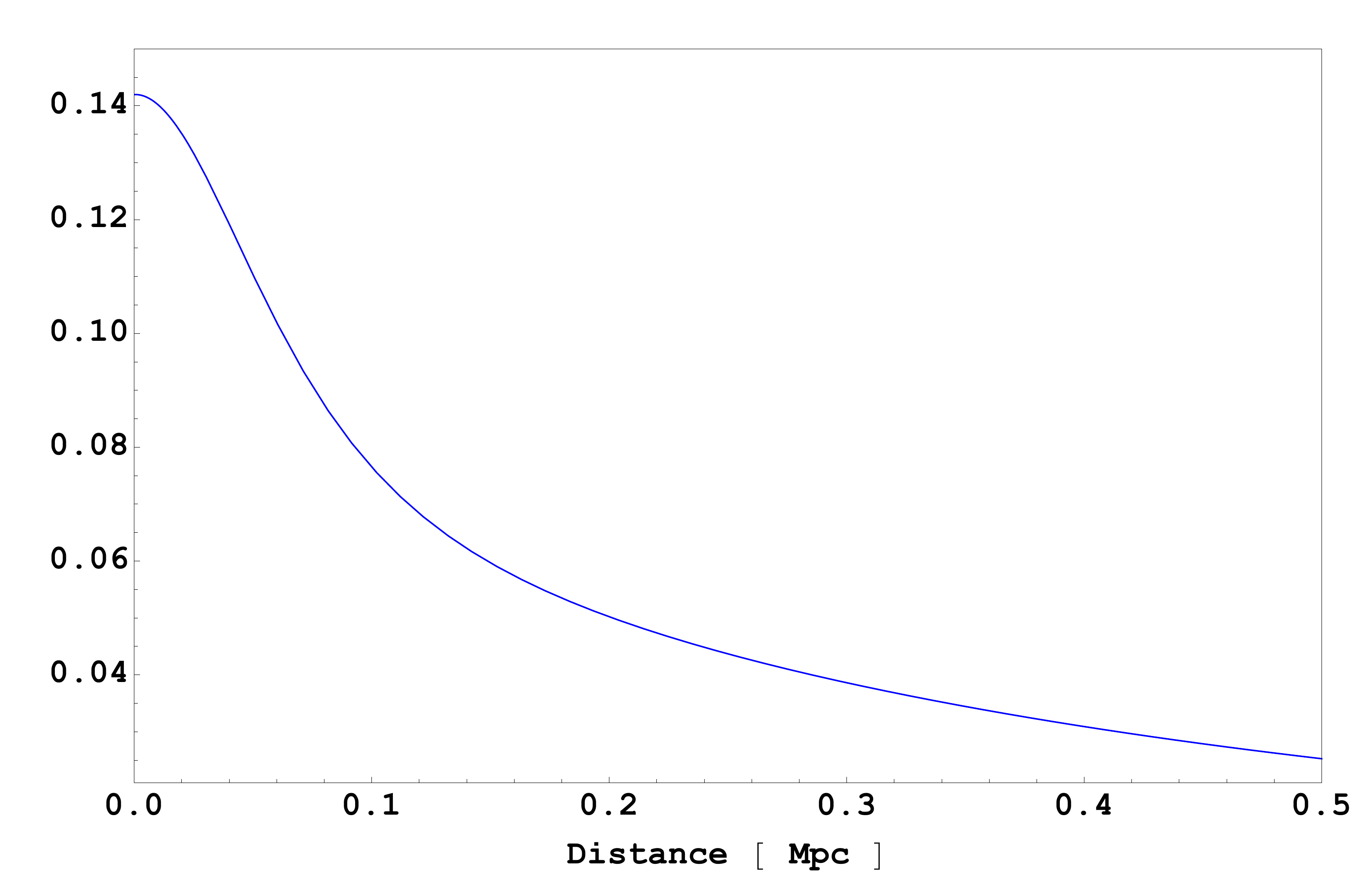}
\caption{Radial profile of the conversion probability $P^{\text{cluster}}_{a\rightarrow \gamma}$ as a function of the distance $R$ from the center 
of Perseus, showing a sharp central peaking behaviour. The probability is given in arbitrary units since we do not specify $M$, see~\eqref{Pcluster}. }
\label{prop_fig}
\end{figure}

As a simple model for the magnetic field in the cluster we take
\begin{equation}
  B(r) = B_0 \sqrt{\frac{n_e(r)}{n_e(0)}} \,,\label{Beta}
\end{equation}
where $B_0  = 25 ~\mu G$ with coherence length 1 kpc and a cluster size of 1 Mpc. This is a deliberately simplistic model compared to the actual 
turbulent, multi-scale cluster magnetic field. Our purpose in using it is to illustrate our key point: 
the photon flux in our scenario exhibits a clear peak in the central region in which the magnetic field is enhanced, see Figure~\ref{prop_fig}.

\subsection{Andromeda}

An X-ray line signal has also been reported in \cite{14024119} for the Andromeda galaxy.
While generally one would expect galaxies to have suppressed signals compared to galaxy clusters, the actual calculation of the
expected flux from the Andromeda galaxy requires the actual magnetic field profile in Andromeda.
As Andromeda is close to edge on, with an inclination angle of 77 degrees, axions emitted from the dark matter halo will have significant passage through the
Andromeda disk, where they can convert to photons in the Andromeda magnetic field. \cite{EgorovPierpaoli} estimate a central magnetic field in Andromeda of $B \sim 50 \mu G$,
and \cite{FletcherBeck2004} report a coherent regular magnetic field of $B \sim 5 \mu G$ between 6 and 14 kpc from the centre of Andromeda.

\subsection{Predictions and Conclusions}

The key feature of this scenario is that the observed photon signal is a convolution of the dark matter density with the
magnetic field structure along the line sight.

Let us enumerate the distinctive predictions this implies:
\begin{enumerate}
\item
A signal strength, or inferred dark matter decay time, that varies from cluster to cluster.
While the position of the line will be identical across clusters (up to the redshift correction),
the strength will vary. Other aspects being equal, clusters with larger magnetic fields will give larger signals.
\item
Within a cluster, the strength of the line will approximately trace the squared magnetic field strength. For cool core clusters with high
central magnetic fields within the core, the line signal should peak at the core (assuming the electron density does not significantly exceed $0.1 ~\hbox{cm}^{-3}$).
\item
If a large stacked sample of clusters is divided into cool-core clusters and non-cool-core clusters, the central region of cool core clusters
should give a stronger signal (i.e. a shorter effective dark matter lifetime) than the central region of non-cool-core clusters.
\item
In environments with high dark matter densities but low magnetic fields, such as dwarf galaxies, the line should be suppressed, with the dominant contribution to
axion-to-photon conversion coming from the magnetic field of the Milky Way. For such local dwarf galaxies, the signal should be stronger for those
closer to the plane of the Milky Way, and also stronger for those at low values of galactic longitude $\vert l \vert$ (so the axions pass through more of the Milky Way
before reaching earth).

\item
For observations of the line from local spiral galaxies, the signal should be stronger for spiral galaxies which are edge on to us compared to those which are face on.
For edge on galaxies, the axions produced by dark matter decay propagate through the disk of their host galaxy, where the magnetic field is largest and with the
largest coherence scales. For face on spiral galaxies, the axions reach us by propagating out of the plane of their host galaxy. This reduces their time
spent in the magnetic field of their host galaxy, reducing the conversion probabilities.
\end{enumerate}
These predictions will be made sharper by better knowledge of the intracluster magnetic fields, as will be provided by for example LOFAR \cite{13053550} and SKA.

\section*{Acknowledgments}

We thank E. Bulbul for clarifying a point in \cite{14022301} and A. Payez, A. Ringwald and M. Zahlen for helpful discussions.
We thank the Isaac Newton Institute for hospitality while finishing the paper. We are supported by the Royal Society and the ERC grant
`Supersymmetry Breaking in String Theory'.


\bibliographystyle{h-physrev}
\bibliography{Perseus}

\begin{thebibliography}{10}

\bibitem{14022301}
E.~{Bulbul} {\em et~al.},
\newblock ArXiv e-prints  (2014), 1402.2301.

\bibitem{14024119}
A.~{Boyarsky}, O.~{Ruchayskiy}, D.~{Iakubovskyi}, and J.~{Franse},
\newblock ArXiv e-prints  (2014), 1402.4119.

\bibitem{Ishida:2014dlp}
H.~Ishida, K.~S. Jeong, and F.~Takahashi,
\newblock (2014), 1402.5837.

\bibitem{Finkbeiner:2014sja}
D.~P. Finkbeiner and N.~Weiner,
\newblock (2014), 1402.6671.

\bibitem{Higaki:2014zua}
T.~Higaki, K.~S. Jeong, and F.~Takahashi,
\newblock (2014), 1402.6965.

\bibitem{Jaeckel:2014qea}
J.~Jaeckel, J.~Redondo, and A.~Ringwald,
\newblock (2014), 1402.7335.

\bibitem{Lee:2014xua}
H.~M. Lee, S.~C. Park, and W.-I. Park,
\newblock (2014), 1403.0865.

\bibitem{Abazajian:2014gza}
K.~N. Abazajian,
\newblock (2014), 1403.0954.

\bibitem{Krall:2014dba}
R.~Krall, M.~Reece, and T.~Roxlo,
\newblock (2014), 1403.1240.

\bibitem{Aisati:2014nda}
C.~E. Aisati, T.~Hambye, and T.~Scarna,
\newblock (2014), 1403.1280.

\bibitem{Kong:2014gea}
K.~Kong, J.-C. Park, and S.~C. Park,
\newblock (2014), 1403.1536.

\bibitem{Hamaguchi:2014sea}
K.~Hamaguchi, M.~Ibe, T.~T. Yanagida, and N.~Yokozaki,
\newblock (2014), 1403.1398.

\bibitem{Frandsen:2014lfa}
M.~Frandsen, F.~Sannino, I.~M. Shoemaker, and O.~Svendsen,
\newblock (2014), 1403.1570.

\bibitem{14031710}
S.~Baek and H.~Okada,
\newblock (2014), 1403.1710.

\bibitem{14031733}
K.~Nakayama, F.~Takahashi, and T.~T. Yanagida,
\newblock (2014), 1403.1733.

\bibitem{14031782}
K.-Y. Choi and O.~Seto,
\newblock (2014), 1403.1782.

\bibitem{hepth0602233}
J.~P. Conlon,
\newblock JHEP {\bf 0605}, 078 (2006), hep-th/0602233.

\bibitem{SvrcekWitten}
P.~Svrcek and E.~Witten,
\newblock JHEP {\bf 0606}, 051 (2006), hep-th/0605206.

\bibitem{12060819}
M.~Cicoli, M.~Goodsell, and A.~Ringwald,
\newblock JHEP {\bf 1210}, 146 (2012), 1206.0819.

\bibitem{12105081}
A.~Ringwald,
\newblock Phys.Dark Univ. {\bf 1}, 116 (2012), 1210.5081.

\bibitem{13025647}
R.~Bähre {\em et~al.},
\newblock JINST {\bf 8}, T09001 (2013), 1302.5647.

\bibitem{14013233}
E.~Armengaud {\em et~al.},
\newblock (2014), 1401.3233.

\bibitem{Burrage}
C.~Burrage, A.-C. Davis, and D.~J. Shaw,
\newblock Phys.Rev.Lett. {\bf 102}, 201101 (2009), 0902.2320.

\bibitem{CAB}
J.~P. Conlon and M.~C.~D. Marsh,
\newblock Phys.Rev.Lett. {\bf 111}, 151301 (2013), 1305.3603.

\bibitem{13123947}
S.~Angus, J.~P. Conlon, M.~C.~D. Marsh, A.~Powell, and L.~T. Witkowski,
\newblock (2013), 1312.3947.

\bibitem{Lieu1996}
R.~{Lieu} {\em et~al.},
\newblock Astrophys. J. {\bf 458}, L5 (1996).

\bibitem{Bonamente2002}
M.~{Bonamente}, R.~{Lieu}, M.~K. {Joy}, and J.~H. {Nevalainen},
\newblock Astrophys. J. {\bf 576}, 688 (2002), astro-ph/0205473.

\bibitem{Durret2008}
F.~{Durret}, J.~S. {Kaastra}, J.~{Nevalainen}, T.~{Ohashi}, and N.~{Werner},
\newblock Space Sci. Rev. {\bf 134}, 51 (2008), 0801.0977.

\bibitem{Lieu1996B}
R.~{Lieu} {\em et~al.},
\newblock Science {\bf 274}, 1335 (1996).

\bibitem{astroph9911001}
S.~{Bowyer}, T.~W. {Bergh{\"o}fer}, and E.~J. {Korpela},
\newblock Astrophys. J. {\bf 526}, 592 (1999), astro-ph/9911001.

\bibitem{Bulbul2009}
M.~{Bonamente}, R.~{Lieu}, and E.~{Bulbul},
\newblock Astrophys. J. {\bf 696}, 1886 (2009), 0903.3067.

\bibitem{Bonafede}
A.~{Bonafede} {\em et~al.},
\newblock \aap {\bf 513}, A30 (2010), 1002.0594.

\bibitem{Gong:2008gi}
Y.~Gong and X.~Chen,
\newblock Phys.Rev. {\bf D77}, 103511 (2008), 0802.2296.

\bibitem{12083562}
M.~Cicoli, J.~P. Conlon, and F.~Quevedo,
\newblock Phys.Rev. {\bf D87}, 043520 (2013), 1208.3562.

\bibitem{12083563}
T.~Higaki and F.~Takahashi,
\newblock JHEP {\bf 1211}, 125 (2012), 1208.3563.

\bibitem{13047987}
T.~Higaki, K.~Nakayama, and F.~Takahashi,
\newblock JHEP {\bf 1307}, 005 (2013), 1304.7987.

\bibitem{13054128}
S.~Angus, J.~P. Conlon, U.~Haisch, and A.~J. Powell,
\newblock JHEP {\bf 1312}, 061 (2013), 1305.4128.

\bibitem{Cicoli:2008va}
M.~Cicoli, J.~P. Conlon, and F.~Quevedo,
\newblock JHEP {\bf 0810}, 105 (2008), 0805.1029.

\bibitem{Linde:1996cx}
A.~D. Linde,
\newblock Phys.Rev. {\bf D53}, 4129 (1996), hep-th/9601083.

\bibitem{Sikivie}
P.~Sikivie,
\newblock Phys.Rev.Lett. {\bf 51}, 1415 (1983).

\bibitem{Raffelt}
G.~{Raffelt} and L.~{Stodolsky},
\newblock Phys. Rev. {\bf D37}, 1237 (1988).

\bibitem{astroph0602622}
G.~B. {Taylor} {\em et~al.},
\newblock \mnras {\bf 368}, 1500 (2006), astro-ph/0602622.

\bibitem{07083264}
G.~B. {Taylor} {\em et~al.},
\newblock \mnras {\bf 382}, 67 (2007), 0708.3264.

\bibitem{VaccaA2199}
V.~{Vacca} {\em et~al.},
\newblock \aap {\bf 540}, A38 (2012), 1201.4119.

\bibitem{GovoniA2255}
F.~{Govoni} {\em et~al.},
\newblock \aap {\bf 460}, 425 (2006), astro-ph/0608433.

\bibitem{VaccaA665}
V.~{Vacca} {\em et~al.},
\newblock \aap {\bf 514}, A71 (2010).

\bibitem{ChurazovForman}
E.~{Churazov}, W.~{Forman}, C.~{Jones}, and H.~{B{\"o}hringer},
\newblock \apj {\bf 590}, 225 (2003), astro-ph/0301482.

\bibitem{08014253}
J.~{Graham}, A.~C. {Fabian}, and J.~S. {Sanders},
\newblock \mnras {\bf 386}, 278 (2008), 0801.4253.

\bibitem{Simi}
A.~Simionescu {\em et~al.},
\newblock Science {\bf 331}, 1576 (2011),
  http://www.sciencemag.org/content/331/6024/1576.full.pdf.

\bibitem{EgorovPierpaoli}
A.~E. {Egorov} and E.~{Pierpaoli},
\newblock \prd {\bf 88}, 023504 (2013), 1304.0517.

\bibitem{FletcherBeck2004}
A.~{Fletcher}, E.~M. {Berkhuijsen}, R.~{Beck}, and A.~{Shukurov},
\newblock \aap {\bf 414}, 53 (2004), astro-ph/0310258.

\bibitem{13053550}
M.~P. {van Haarlem} {\em et~al.},
\newblock \aap {\bf 556}, A2 (2013), 1305.3550.

\end{thebibliography}

\end{document}